\documentclass{svproc}
\usepackage{url}
\usepackage{bbm,bm,amsmath,amssymb}

\begin{document}
\mainmatter              
\title{de Sitter Vacua in the String landscape:\\ La Petite Version}
\titlerunning{de Sitter}  
%
\author{Keshav Dasgupta\inst{1} \and Maxim Emelin\inst{1} \and
Mir Mehedi Faruk\inst{1} \and Radu Tatar\inst{2}}
\authorrunning{Keshav Dasgupta et al.} 
%
\tocauthor{Keshav Dasgupta, Maxim Emelin, Mir Mehedi Faruk and Radu Tatar}
\institute{Department of Physics, McGill University, Montreal QC H3A 2T8, Canada\\
\email{keshav@hep.physics.mcgill.ca; maxim.emelin, mir.faruk@mail.mcgill.ca}\\ 
\and
Department of Mathematics, University of Liverpool, Liverpool L69 7Zl, UK\\
\email{Radu.Tatar@Liverpool.ac.uk}}

\maketitle              

\begin{abstract}
In this review we argue that  four-dimensional effective field theory descriptions with de Sitter isometries 
are allowed in the presence of time-dependent internal degrees of freedom in type IIB string landscape. Both moduli stabilizations and time-independent Newton constants are possible in such backgrounds. However once the time-dependences are switched off, there appear no possibilities of  effective field theory descriptions and  these backgrounds are in the swampland.
\end{abstract}
\section{Introduction}
The late time (quasi-)de Sitter vacuum is a desired consequence of any UV complete theory of quantum gravity, therefore one would expect its appearance in string theory. Earlier attempts to reproduce such a vacuum 
in the context of type II theories \cite{KKLT} have had some successes, although more recently various questions have been raised (see for example \cite{bena} \cite{sav} \cite{westphal}) regarding the validity of 
such computations. The answers to some of these objections are attempted in  \cite{soler} and 
\cite{kalloshevan}, mostly by clarifying the role of anti D3-branes as one of the key ingredients in realizing a four-dimensional de Sitter vacuum.
A more serious objection, not just against the computations of \cite{KKLT}, but against the very existence of de Sitter vacua in the string landscape, has been recently raised in \cite{vafa} claiming that string theory prefers quintessence  instead of an inflationary evolution towards a late time de Sitter vacuum. The situation is rather paradoxical because
objections against models of quintessence were around from early on, and more recently \cite{HB} showed that the swampland criteria of \cite{vafa} are in fact {\it inconsistent} with simple models of quintessence. More general arguments against swampland criteria themselves have also been raised mostly concerning their implications on cosmology (see for example \cite{clovB}), or their implications on general effective field theories \cite{kochu}.  The latter paper in fact questioned the adhoc nature of the criteria themselves. As a response to these objections,  in 
\cite{vafa2}, the criteria were given a slightly formal derivation using the trans-Planckian problems first raised for inflationary cosmology in \cite{transplanckian}. The original swampland criteria of \cite{vafa} are then elevated to a 
Trans-Planckian Censorship Conjecture (TCC), meaning that the validity of the criteria \cite{vafa} rely on censoring the trans-Planckian modes in a theory of quantum gravity because these modes have non-unitary evolutions. Whether this is really a problem in string theory, where we expect a well defined behavior at short distances, remains to be seen (see \cite{yamazaki} for a discussion on related aspect). 

In this review, which basically summarizes some parts of \cite{deft} and \cite{deft2}, we will argue that if the internal degrees of freedom in a type IIB compactification with four-dimensional de Sitter isometries, remain time independent, then there is no four-dimensional effective field theory description possible. Such backgrounds may truly be in the swampland, thus providing some credence to \cite{vafa}. However once the internal degrees of freedom become time-dependent (but preserving the four-dimensional Newton constant), then an effective field theory description becomes possible.

\section{How hard is to get a de Sitter vacuum in string theory?}
Let us consider two sets of backgrounds: one, in which we allow a four-dimensional de Sitter space with a flat slicing and an internal six-dimensional compact space with time-independent warp-factors; and two, a similar compactification but now with an internal six-dimensional space with time-dependent warp-factors. Additionally, for the first case we allow all internal degrees of freedom to be time-independent (the internal degrees of freedom being the three and five-form fluxes; as well as the axio-dilaton), whereas for the second case we 
allow all internal degrees of freedom to be time-dependent. In both cases however we keep vanishing axion but constant dilaton, and time-independent Newton's constant. Question is, which of these two backgrounds would solve the equations of motion in type IIB theory? 

In the absence of quantum corrections, even if we allow fluxes, branes, anti-branes and/or orientifold planes, neither of the two backgrounds can solve the sugra equations of motion as was shown in 
\cite{nogo} and \cite{nodS}. This is of course a consequence of the no-go theorem first proposed in 
\cite{giburam}. The situation becomes more interesting when quantum corrections are added in. The naive expectation would be that  once a sufficient number of quantum terms are added in, either of the two background should solve the equations of motion. Unfortunately this expectation didn't quite turn out to be correct as was first demonstrated in \cite{nodS} and \cite{deft}. In the absence of time-dependent internal degrees of freedom, the equations of motion can {\it never} be solved because the quantum corrections do not lead to an effective field theory description in four-dimensions with de Sitter isometries 
\cite{nodS} \cite{deft} \cite{deft2}.
Thus these backgrounds are truly in the {\it swampland} \cite{vafa}. On the other hand, once the internal degrees of freedom become time-dependent, miraculously the effective field theory description 
reappears \cite{deft} \cite{deft2}. 

Why is this the case? The answer can be ascertained from many angles, but we shall adopt the strategy of uplifting the IIB background to M-theory and then studying the dynamics from there. The reason for choosing M-theory as opposed to IIB is not just for the sheer compactness of the representation of the degrees of freedom (the {\it number} of degrees of freedom of course does not change from IIB to M-theory), but also for the fact that M-theory allows a Lagrangian formulation (even with non-local counter-terms, as we shall discuss a bit later) as opposed to IIB, where a Lagrangian formulation is harder to come by. Existence of a Lagrangian formulation then allows us to express the most generic form of the quantum corrections as a series in polynomial powers of four-form flux components and Riemann curvature tensors as well as with spatial and temporal derivatives as shown in   
eq. (2.6) of \cite{deft2}. The spatial derivatives are with respect to the internal eight-manifold which is an orbifolded torus fibered over the six-dimensional base that we have in the IIB side. The existence of an effective field theory in lower dimensions then depends on whether the quantum series allows a hierarchy or not. This hierarchy should of course be with respect to $M_p$, but also with respect to any other relevant expansion parameter. The only other allowed expansion parameter is the type IIA string coupling $g_s$, so the question of hierarchy boils down to finding  a {\it finite} number of operators at any order in 
$g_s^{|a|}/M_p^{b}$, where $(a, b) \in \left({\mathbb{Z}\over 3}, \mathbb{Z}\right)$, and the $1/3$ moding has been explained in \cite{deft} \cite{deft2}. Note that while  $b$ can have any sign, we take only positive powers of $g_s$. This is because all the negative powers of $g_s$ can be resummed into powers of 
${\rm exp} \left(-\frac{1}{ g_s^{1/3}} \right)$, 
taking the expected non-perturbative form and expressing the full set of corrections as a resurgent trans-series.

All of these would make sense when $g_s \ll 1$. As a happy coincidence, the type IIA coupling $g_s$ turns out to be related to the dimensionless temporal coordinate in the IIB side (the dimension being determined by the cosmological constant $\Lambda$). This means at late time, where we expect $\Lambda \vert t\vert^2$ to vanish (recall that we are in the flat slicing), the type IIA string coupling $g_s$ also vanishes, implying that the weak coupling limit is also the late time limit in our case. This is good, but the fact that $g_s$ now becomes time dependent implies that one needs to deal with the quantum series (eq (2.6) in \cite{deft2}) much more carefully. In fact one would also have to assign certain temporal dependences of the background G-flux components in M-theory. A simple ansatze for the G-flux is given in eq. (2.5) of \cite{deft2} where, note that, we have traded the temporal dependences with $g_s$ dependences. This means generically all degrees of freedom may be expanded into a series of perturbative as well as non-perturbative $g_s$-dependent terms. At late time, the non-perturbative terms decouple, so we are left with a perturbative series in $g_s$. 

A careful study of the quantum series then reveals the following interesting fact. One can easily isolate the dominant $g_s$ coefficient of the quantum series, which we henceforth express as $g_s^{\theta_k}$, with 
$\theta_k$ is as given by eq. (2.10) in \cite{deft2}. This is expressed in terms of polynomial powers of the
curvature tensor and G-flux components (which are denoted in eq (2.10) in \cite{deft2} by $l_i \in \mathbb{Z}$); parameters $k_i \in {\mathbb{Z}\over 2}$ that denote how the $g_s$ dependences of the various G-flux components are expressed in eq. (2.5) of \cite{deft2}; as well as on $n_0$ and $n_{1, 2, 3}$, the former being related to the number of temporal derivatives and the latter to the number of internal spatial derivatives in the quantum series given as eq. (2.6) of \cite{deft2}.
 If the G-flux components are time independent (i.e when $k_i = 0$), then $\theta_k$, which we write as $\theta_0$, is as in eq. (2.11) of \cite{deft2}.
 
The important thing to note now is the difference between eq. (2.10) and eq. (2.11) in \cite{deft2}. If $k_1 \ge 0$, $k_2 \ge {3\over 2}$ and $n_3 = 0$, then eq. (2.10) has no relative minus signs whereas eq. (2.11) does have relative minus signs. Interestingly switching on $n_3$, i.e derivatives along the toroidal directions, both the expressions have relative minus signs. For the time being let us assume that we have switched off $n_3$ (we will come back to non-zero $n_3$ just a bit later). In that case, the $g_s$ scalings of the energy momentum tensors along the various directions in M-theory (i.e along the space-time, internal six-manifold, and the internal toroidal directions) now become eq. (2.12) of \cite{deft2}. What does that imply? 

Let us take the $2+1$ dimensional space-time directions to clarify the implications of the above $g_s$ scalings. The energy-momentum tensors of the quantum terms will appear on the RHS of the Einstein's equation. The $g_s$ scalings of parts of the Einstein's equation can be shown to scale as $g_s^0$, i.e they are at zeroth order in $g_s$ (there is some subtlety associated with the scaling as shown in section 4.1 of \cite{deft}). Now comparing the $g_s$ scalings of the quantum terms along the space-time directions, i.e eq. (2.12) of \cite{deft2}, we see that $\theta_k = {8\over 3}$, implying that the quantum terms are constructed out of eighth order polynomials in curvature and G-flux components. This seems like a marvellous thing, until we notice that for the time-independent case the equations $\theta_0 = 
{8\over 3}$ do not have a {\it finite} number of integer solutions for ($l_i, n_{1, 2}, n_0$). The loss of finiteness is precisely because of the presence of relative {\it minus} signs in the expression for $\theta_0$ in eq. (2.12) of \cite{deft2}. On the other hand, with $\theta_k = {8\over 3}$ in eq. (2.11), there are no relative minus signs (we have made $n_3 = 0$), so there is only a {\it finite} number of operators to zeroth order in $g_s$.  

Such a conclusion would appear to immediately rule out the time-independent compactifications, because allowing an infinite number of operators means that we have lost $g_s$ hierarchy.  However the scenario is  a bit more subtle than what appears at face-value. In fact a careful analysis reveals a few caveats. First is of course the case with $n_3 > 0$. Second, and this may be more important, all the quantum terms (finite or infinite) at zeroth order in $g_s$ have {\it different} $M_p$ suppressions. Thus although we have lost our $g_s$ hierarchy for the time-independent case, we seem to still retain an $M_p$ hierarchy in the system. Could we then just make $M_p \to \infty$ and get rid of operators suppressed by higher powers of $M_p$ and retain a finite set of operators for both time-dependent and time-independent cases respectively? What goes wrong with such a line of reasoning?

The answer lies in the careful mapping between the degrees of freedom of IIB in M-theory. For example not all G-flux components can be allowed without breaking the four-dimensional de Sitter isometries in the IIB side. In particular, if we allow the internal G-flux components with two components along the toroidal directions, they cannot be global fluxes. In fact they can only be localized fluxes so that they appear as gauge fluxes on seven-branes in the IIB side (the seven branes being related to the orbifold points in M-theory, and are arranged as in \cite{senmukhi} so that vanishing axion and constant dilaton may still be maintained). In the simplest case, the localized pieces of the G-flux components can be viewed as a gaussian over the toroidal space. The spread of such gaussian pieces is measured in terms of $M_p$, at least for the time-independent cases, implying that the localized fluxes could in principle have hidden $M_p$ scalings implicit in their definitions themselves. Such $M_p$ scalings are in general harmless because they do not interfere or change the $M_p$ suppressions of the quantum terms. However the quantum terms, i.e eq. (2.6) of \cite{deft2}, do also have derivatives along the toroidal directions which we called $n_3$ earlier. If we switch on these $n_3$ derivatives, we see that now they are able to influence the $M_p$ scalings of the quantum terms!    

As shown in \cite{deft2}, the $n_3$ derivatives change the $M_p$ scalings of the quantum terms in an interesting way: they introduce a relative minus sign there. The $g_s$ scaling $\theta_0$ already had their share of minus signs, and now putting everything together we can ask, for the time-independent case, how many operators are allowed at zeroth order in $g_s$ and $M_p$ (we can even ask at zeroth order in $g_s$ but second order in $M_p$).  The answer is as shown in eq. (2.23), defined in terms of certain variables that may be expressed in terms of ($l_i, n_i, n_0$) in eq. (2.16), both equations being from \cite{deft2}. The worrisome thing is the relative {\it minus} signs in eq. (2.23) whose RHS is 8, related to the eighth order polynomials in curvature and G-flux components. There are an {\it infinite} number of integer solutions to this equation, all to zeroth order in $g_s$ and $M_p$. Now we have clearly lost both the $g_s$ and $M_p$ hierarchies, implying that there {\it cannot} be an effective-field theory description in the IIB side with de Sitter isometries and with time-independent internal degrees of freedom.

The above conclusion is definitely intriguing, but couldn't we say the same thing for the time-dependent cases too when we switch on the $n_3$ derivatives? Why should the time-dependent cases be any different? Answering this will take us to the second level of subtleties in our construction, that we hitherto kept under the rug.

In the time-dependent case, most of the arguments related to the localized fluxes go through in a similar way to the time-independent case. However there are now a few subtle differences. The first difference lies in the scale of the gaussian spread of the localized function. As shown in eq. (2.25) in \cite{deft2}, the gaussian spread now involves {\it both} $g_s$ and $M_p$, simply because the distances are measured in terms of {\it warped} metric components (that involve $g_s$ dependences). This means the $n_3$ derivatives in the quantum terms of 
eq. (2.6) will contribute to both $g_s$ and $M_p$ scalings. We already discussed how the $M_p$ scalings change, so the question now is whether the change in the $g_s$ scalings can alter our earlier conclusion. The answer turns out to be a bit more subtle. If the G-flux ansatze for the localized flux, as given by eq. (2.15) and eq (2.25) of \cite{deft2}, has $g_s$ independent coefficients (i.e ${\cal A}_{n0}$ in eq. (2.25)), then even the extra $g_s$ contributions from $n_3$ derivative cannot change the conclusion. The reason is simple to state: with time-independent coefficients we are as though allowing $g_s$ independent pieces in our original G-flux ansatze eq.  (2.5) of \cite{deft2}. This would be like having $k_2 = 0$ in eq (2.6) of \cite{deft2}, leading to the above-mentioned problems. Thus the coefficients of the localized G-flux components should also have $g_s$ dependent factors. Such factors easily arise by simply normalizing the localized functions!
Putting everything together immediate reproduces eq. (2.28) of \cite{deft2}, implying {\it finite} number of 
operators at zeroth orders in $g_s$ and $M_p$ with time-dependent internal degrees of freedom. 

The story doesn't end here because M-theory could also have non-local quantum terms, these terms are sometime christened as non-local counter-terms. A simple way to infer their presence is to go to the infinite derivative limits of eq. (2.6) in \cite{deft2}. More generic form of non-localities are possible in the limit where 
($n_0, n_i$) themselves become {\it negative}. Such negative-derivative actions may be rewritten as nested 
integrals as shown in \cite{deft} and \cite{deft2}, thus allowing a Lagrangian formulation of the scenario. One however needs to be careful in introducing such terms because we don't expect the non-localities to show up in the low energy limit of M-theory. The validity of the non-local quantum terms has been discussed in great details in \cite{deft}, and we urge the readers to look up the discussions therein.  What we want to question here is what happens to the two cases once we switch on these non-local quantum terms.    

The answer as meticulously shown in section 3 of \cite{deft2}, to any given order in 
$g_s^{|a|}/M_p^b$, there are only finite number of operators possible when all the internal degrees of freedom become time-dependent, whereas the number of such operators tend to become infinite once time-dependences are switched off. These have been demonstrated in figure 1 to 4 in \cite{deft2}, where moving vertically down in any figure counts the number of operators. The vertical axis in each figure represents the level of non-locality (i.e how many nested integrals are taken into account), and the horizontal axis represents the $M_p$ scalings. Thus it appears that non-localities do not seem to change the generic idea that four-dimensional effective field theory description with de Sitter isometries remains valid whenever the internal degrees of freedom become time-dependent, but fails when the time-dependences are switched off.
Interestingly all of these happen while still keeping the four-dimensional Newton constant time-independent.

There are many other questions that may be asked at this stage, for example: How are the moduli stabilized? How are the equations of motion satisfied? How do the flux quantization, and anomaly cancellation work out in the time-dependent cases? All of these have been answered, and to preserve the brevity of the note we will refer the interested readers to our longer works \cite{deft} and \cite{deft2}. We will however make two comments, one, related to the quantum potential as it appears in eq. (2.7) of \cite{deft2} and two, related to the possibility of realizing inflation in our construction. 

The quantum potential, with contributions from both local and non-local quantum terms, gives rise to an exact expression for the renormalized cosmological constant $\Lambda$ as given in eq. (4.1) of \cite{deft2}. This contribution appears from the zeroth order in $g_s$ and second order in $M_p$ (i.e to order $g_s^0/M_p^2$ once we assume the non-local contributions are suppressed by the toroidal volume in M-theory). Why 
don't the higher order quantum terms contribute to the cosmological constant? The answer as detailed in 
\cite{deft} and \cite{deft2} is easy to see: at higher orders in $g_s$ we also switch on other G-flux components  
 that appear at higher orders in $k$ in eq. (2.5) of \cite{deft2}. These components back-react on the geometry creating {\it negative} gravitational potentials. These negative gravitational potential are then exactly cancelled by the positive contributions from the quantum potential at higher orders in $g_s$ in such a way that the zeroth order cosmological constant  remains unaltered. 
 
The final comment is on the possibility of realizing inflation on our set-up. It has been claimed in \cite{vahid} 
that warm inflation may indeed be outside the swampland. Here however we want to comment on the possibility of using the time-dependences to allow for some variant of the D3-D7 inflationary model of 
\cite{DHHK}, the D7-branes in IIB appearing from the orbifold points in M-theory and the D3-branes appearing from the M2-branes that we require the cancel anomalies in the system. This of course calls for a F-theory uplift of our IIB picture (as in \cite{senmukhi}), so a natural question is to see whether the M-theory and F-theory connection could be made more precise with time-dependent internal degrees of freedom. It would also be interesting to compare our answers with those of \cite{ding}.  

\section{Discussion and conclusion}
In this short note we have summarized some of the contents of \cite{deft} and \cite{deft2} related to the possibility of the existence of four-dimensional effective field theory description with de Sitter isometries 
and time-dependent internal degrees of freedom in IIB. However there are still questions related to the 
exact meaning of Wilsonian effective actions in such spaces, for example how the time-dependent degrees of freedom are integrated out, and whether the de Sitter solutions that we have could be interpreted as {\it vacua} or coherent states over some supersymmetric solitonic minima.  Additionally, the precise connection to TCC needs to be spelled out, maybe along the lines of the recent work \cite{brahma} (see also \cite{yamazaki}). All these and other advances show that this is indeed an exciting time for cosmology.  

%
%

\end{document}